# Hard magnet coercivity

S. BANCE[1], G. CIUTA[2,3], T. SHOJI[4], T. GAO[3], G. HRKAC[5], M. YANO[4], A. MANABE[4],
N. M. DEMPSEY[2,3], T. SCHREFL[1,6], D. GIVORD[2,3,7]

[1] Dep. of Technology, St Pölten University of Applied Sciences, Austria
[2] Univ. Grenoble Alpes, Inst NEEL, F-38042 Grenoble, France
[3] CNRS, Inst NEEL, F-38042 Grenoble, France
[4] Toyota Motor Corp., Toyota City, Japan
[5] CEMPS, University of Exeter, Exeter, UK
[6] Center for Integrated Sensor Systems, Danube University Krems, Austria
[7] Instituto de Fisica, Universidade Federal do Rio de Janeiro, Rio de Janeiro RJ, Brasil

**Abstract:** *Based on a critical analysis of the experimental coercive properties, general considerations on the reversal mechanisms in RFeB magnets are recalled. By plotting together the experimental parameters obtained in various magnets, common features of the reversal processes are demonstrated. Modeling provides an almost quantitative description of coercivity in these materials and permits connecting the defect characteristic properties to reversal mechanisms.*

**Address:** Dominique Givord, Institut Néel, CNRS/UJF, BP166 38042 Grenoble-cedex 9 France, +33476881090 e-mail: dominique.givord@neel.cnrs.fr

## 1. Introduction

The understanding of coercive phenomena is one of the most challenging aspects of the study of hard magnetic materials [1,2]. As well known, the reversal process of reference is coherent rotation, as described by the Stoner-Wohlfarth model and the associated coercive field, $H_c$, is equal to the anisotropy field, $H_A$ [3]. The real coercive field is significantly smaller than the anisotropy field and this is ascribed to the influence of defects. The reversal process involves the formation of transitory heterogeneous magnetic states. Schematically, these can be listed as: nucleation of a region of non-saturated magnetization (with respect to the initial magnetic state), propagation/expansion and (possibly) pinning [2]. The study of reversal processes aims at identifying the one that is critical, i.e. governs full reversal within the individual objects (grains) from which the magnet is made. Reversal phenomena are not accessible to direct experimental observation and their analyses rely on indirect measurements. These are the temperature dependence of the coercive field, the angular dependence of the coercive field and the temperature dependence of the activation volume. In the first sections of this paper, the conclusions that can be drawn from such measurements are discussed. This part of the manuscript is not original but it is necessary to understand the significance of the second part, in which common experimental features of the reversal mechanisms are revealed and the results of numerical modeling are presented. Modeling provides an almost quantitative description of coercivity and permits connecting the defect characteristic properties to reversal mechanisms.

## 2. Physical significance of the temperature dependence of the coercive field

In the analysis of coercivity, it is common to compare the temperature dependence of the coercive field to that of intrinsic physical parameters characterizing the hard magnetic phase. On purely phenomenological grounds, Kools introduced the expression [4] :

$$\mu_0 H_c = \alpha_m \mu_0 H_A - \mu_0 H_D, \qquad (1a)$$

where $H_A$ is the hard phase anisotropy field and $H_D = -N_{eff} M_s$ is the demagnetizing field ($M_s$ is the spontaneous magnetization), $\alpha_m$ and $N_{eff}$ are adjustable parameters. The evaluation of the strength of the demagnetizing field (i.e. of the parameters $N_{eff}$) in such hard magnetic materials is a difficult task [1]. Due to the heterogeneous character of the magnetic configuration, the divergence of the magnetization is not zero. In addition to the usual surface charges, volume charges contribute to the demagnetizing field and their value depend on the reversal processes themselves. However, the demagnetizing field is not expected to modify qualitatively the nature of the reversal mechanisms. Since the focus of the present paper is on the discussion of such mechanisms, $N_{eff}$ is considered here as a purely phenomenological parameter. Expression (1a) can be re-written as:

$$\mu_0 H_T = \alpha_m \mu_0 H_A, \qquad (1b)$$

where $\mu_0 H_T = \mu_0 H_c - \mu_0 H_D$ is the total field determining reversal, obtained by subtracting $H_D$ to the experimental coercive field $H_c$ ($|H_T| > |H_c|$).

Within the so-called micromagnetic model, developed by Kronmüller and co-workers [1,6], a planar defect is assumed. The possible reversal mechanisms are coherent rotation-like or domain-wall de-pinning. The nature of the dominant mechanism may be derived from the value of the parameter $\alpha_m$ in expressions (1). In the case of the NdFeB magnets, its value, is of about $1/3$. Within the present model, this is incompatible with pinning. This led to conclude that the reversal process is coherent rotation-like [1,5].

The other expression to which the temperature dependence of the field $H_T$ may be compared is obtained within the so-called global model [6, 7]. This model considers that the various possible reversal mechanisms, except coherent rotation, involve the formation of a heterogeneous magnetic configuration. This must be somehow similar to a domain wall, the magnetic configuration of minimum energy that incorporates magnetization reversal. This gives:

$$\mu_0 H_T = \alpha_G \frac{\gamma}{v^{1/3} M_s} \qquad (2)$$

where $\gamma$ is the domain wall energy within the hard phase and $v$, the activation volume, an experimental parameter. Expression (2) is expected to account for the changes in both the domain wall energy and the domain wall surface area that may occur during reversal, i.e. it may represent the various processes described in the introduction.

The present discussion suggests that the nature of the reversal mechanisms may be identified, at least partially, from the analysis of the temperature dependence of the coercive field, More precisely, the applicability of expression (2) and not that of expression (1) should exclude that reversal occurs by coherent rotation. However, it turns out that this argument does not apply. Both expressions (1) and (2) generally provide satisfactory account for the experimental temperature dependence of $H_c$ [2, 8]. This is due to the fact that the experimental value of $v^{1/3}$ is approximately proportional to the domain wall width, $\delta$ (see next section) [2]. Most hard magnetic materials crystallize within a uniaxial structure for which the magnetocrystalline anisotropy is generally dominated by second-order terms. In this case, the domain wall thickness $\delta = \pi\sqrt{A/K}$ and $\gamma = 4\sqrt{AK}$, leading to $\gamma/v^{1/3} \sim \gamma/\delta \sim K$ ($A$ is the exchange constant and $K$ the second order magnetocrystalline anisotropy coefficient). Since relations (1) and (2) are formally equivalent, they cannot be used to distinguish between reversal mechanisms.

However, this argument does not apply to NdFeB magnets below typically 200 K. In this temperature range, higher order anisotropy terms become important in $Nd_2Fe_{14}B$, leading to the spin reorientation that takes place at 135 K. It is then found that the global model tends to provide better account for the temperature dependence of $H_c$ than the micromagnetic model (see [2]).

At the end of this section, it should be stressed that the applicability of both expressions (1) and (2) implies that the coercive field $H_c$ is proportional to some intrinsic parameters characterizing the hard magnetic phase. Assuming that reversal occurs by coherent rotation, then, $H_c \ll H_A$ implies that $H_{A'} \ll H_A$, whereby $H_A'$ is the anisotropy field in the defect region. From one material to another, $H_A(T)$ may vary considerably, and it is puzzling that $H_{A'}$ should remain proportional to $H_A$. Amongst others, this experimental observation led us to conclude that reversal at $H_c$, is governed by the passage/expansion of a pre-formed nucleus [7].

## 3. Temperature dependence of the activation volume

The concept of the activation volume is intrinsically linked to that of coercivity [6, 9-11]. Magnetization reversal occurs when the difference between the minimum energy configuration and the saddle point energy configuration can be provided by thermal activation. The characteristic time for magnetic measurements is of the order of 10-100 s, leading to the classical thermal energy value of $25 k_B T$. Experimentally, the activation volume is obtained by comparing the field dependence and the time dependence of the magnetization. At "low temperature " (up to room temperature for NdFeB magnets), $v$ remains approximately proportional to $\delta^3$ [2,11], i.e. the activation volume magnetic properties are relatively close to the main phase properties, in agreement with previous section. At higher temperature, the activation volume increases much faster with T than $\delta^3$. This suggests that the anisotropy in the activation volume decreases faster with T than its main phase value. The anisotropy is approximately proportional to the magnetization at some power (anisotropy coefficients of order $l$ should vary as $M_s^{l(l+1)/2}$ according to the Akulov law), suggesting that the magnetization in the activation volume decreases faster than bulk magnetization with increasing temperature. Numerical modeling (see below) reveals that defects that are less than 1 nm thick, already affect considerably the value of the coercive field. The high temperature dependence of $v$ may thus be tentatively related to the temperature dependence of the magnetization being higher at the surface than in the bulk.

**4. Angular dependence of the coercive field**

The angular dependence of the coercive field has been discussed in previous publications [12-14]. It is closer to Kondorsky's $1/\cos\theta$ law than to Stoner-Wohlfarth's law. This is another result suggesting that magnetization reversal does not occur by coherent rotation but involves the formation of a domain wall-like configuration, so that low coercive field can be reconciled with an anisotropy in the activation volume of the same order as main phase anisotropy.

**5. Comparing the room temperature coercivity in various NdFeB magnets**

At this stage, the activation volume and the coefficient $\alpha_G$ are parameters that permit describing the temperature dependence of coercivity, but to which no special physical significance is attached. To go beyond this, the properties of different samples and the associated parameter values should be compared. Equation (2) suggests plotting $H_T$ as a function of $1/v^{1/3}$. This is done in Fig 1 for various RFeB magnets. $H_T$ increases as the size of the activation volume decreases, as expected. However, the approximate linear variation seen in Fig. 1, does not extrapolate to the origin of the coordinates, as it would do for constant $\alpha_G$. The slope of the variation, higher than expected, tells us that the parameter $\alpha_G$ increases with the coercive field.

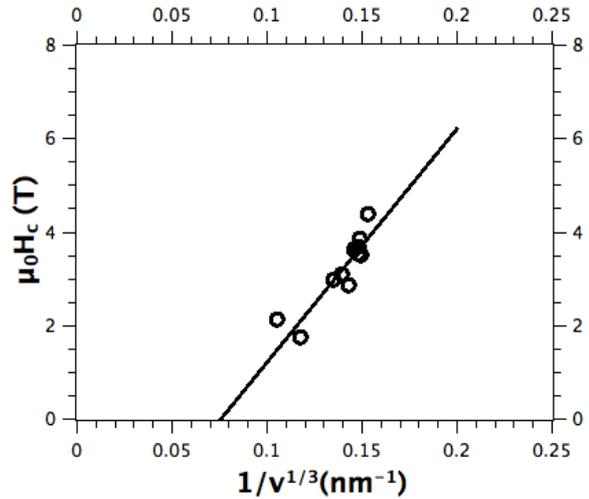

*Fig. 1 : Coercive field, $\mu_0H_c$ versus $1/v^{1/3}$ ($v$ = activation volume) for a series of RFeB magnets.*

Coming back to expression (2), we will now assume that the magnetocrystalline anisotropy is determined by second order terms only. The domain wall energy and the domain wall width in $v$, are given by $\gamma' = 4\sqrt{A'K'}$ and $\delta' = \pi\sqrt{A'/K'}$ where $A'$ is the exchange constant in the activation volume. Assuming the activation volume proportional to $\delta'^3$ leads to [7]:

$$\mu_0 H_T = \alpha_G' \frac{4\pi A}{v^{2/3} M_s} \qquad (3)$$

Expression (3) presents a fundamental difference with expression (2) above. No a priori assumption on the value of the anisotropy in the activation volume is made. The exchange constant, $A'$, is assumed proportional to the exchange constant, $A$, within the hard phase. This is a much less stringent hypothesis than the usual assumption $K'$ proportional to $K$, since anisotropy is by far more sensitive to small structural defects than exchange interactions are.

Expression (3) may account for $H_c(T)$ in ferrite and Pr-Fe-B magnets [7]. It cannot be used however to describe $H_c(T)$ in NdFeB magnets for which a spin-reorientation occurs at 135 K, i.e. anisotropy constants of higher orders are not negligible and even larger than the second order anisotropy constant in a certain temperature range. However, at room temperature, relation (3) should apply. The field $\mu_0 H_T$ is plotted as a function of $1/v^{2/3}$ in Fig. 2 for the same magnets as in Fig. 1. An approximate linear relationship is found and the corresponding line extrapolates to the origin of the coordinates. This shows that the coefficient $\alpha'_G$ has approximately the same value in all samples. The slope derived from figure 2 amounts to 160 T/nm². Using $A = 8 \cdot 10^{-12}$ J/m and $\mu_0 M_s = 1.61$ T, gives $\alpha'_G = 2$.

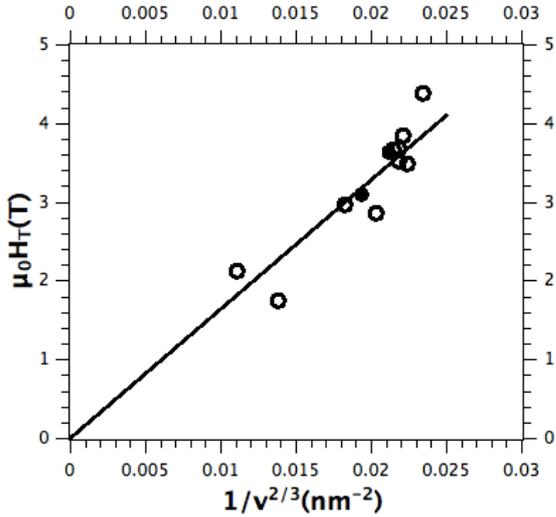

Figure 2 : Coercive field, $\mu_0 H_c$, versus $1/v^{2/3}$ (v is the activation volume) (same samples as Fig. 1). The line is a fit to the data, with slope 160 T/nm².

The constancy of the parameter $\alpha'_G$ implies that the fundamental geometry of the reversal process remains essentially unchanged as the coercive field is varied within a factor of almost 3. Assuming $A'=A$, a simple expression for $\alpha_G$, is obtained by comparing expressions (2) and (3) :

$$\alpha_G = \frac{\alpha'_G}{v^{1/3}} \delta = 2\frac{\delta}{v^{1/3}} \qquad . \qquad (4)$$

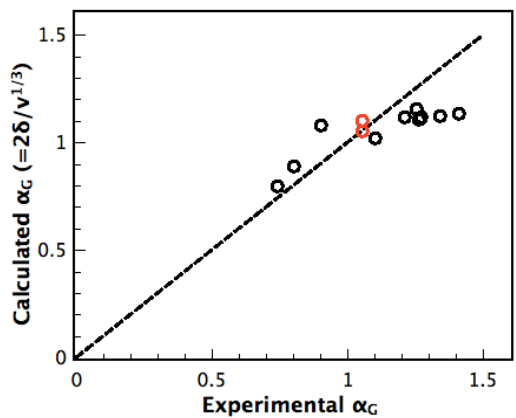

Figure 3 : Calculated $\alpha_G$ (= $2\delta/v^{1/3}$) compared to experimental $\alpha_G$ derived from the temperature dependence of the coercive field; $\delta$ is the $R_2Fe_{14}B$ domain wall width and v the activation volume.

The increase of $\alpha_G$ with $H_T$ (Fig. 1) is thus an illustration of the inverse dependence between $H_T$ and v ( [2] and [7]). The coefficient $\alpha_G$ given by expression (4) is plotted in Figure 3 as a function of the experimental $\alpha_G$, derived from the analysis of the temperature dependence of $H_c$. A striking correspondence is found between both coefficients, despite the fact that they were derived using two different methods and that the calculation of $\alpha_G$, involves a unique adjustable parameter, $\alpha'_G$ of which value is the same for all samples. These results demonstrate the global consistency of the present analysis.

**Numerical modeling**

The analysis described in the above section provides a consistent description of coercivity, within a certain physical representation of the reversal process. In the present section, the reversal processes are examined in the framework of recent numerical modelling studies [15]. The polyhedral hard magnetic grains were approximated by simple cubes with a defect shell of zero anisotropy. The magnetization reversal process was computed numerically by solving the equation of motion for the magnetization, with temperature-dependent intrinsic material paramaters. As opposed to linearized micromagnetic models, both the linear and non-linear nature of the equations were taken into account. It was found that magnetization reversal begins at a corner of the cube, where increased demagnetizing field causes localized curling of the magnetization.

As expected, in the absence of any defect, the coercive field ($\mu_0 H_c \sim 5.8\ T$) approaches the value for coherent rotation ($H_c=H_A$), once the demagnetizing field contribution is considered ($\mu_0 H_D \sim \mu_0 M_s \sim 1.6T$). However, as noted above, even in this case, reversal does not occur by coherent rotation. Due to the heterogeneous character of the demagnetizing field, a magnetic configuration forms that appears to resemble a domain wall. It is generally accepted that the existence of defects is almost unavoidable, of which presence might actually be intrinsic to the material surface [17]. The existence of defects is simply taken into account by assuming the presence of a zero anisotropy layer at the grain surface. As the thickness of the defect layer, $t$, is increased up to 2 nm, the 300 K coercive field decreases from 5.8 T to 1.6 T. Most importantly, as soon as the defect thickness exceeds 0.5 nm, a stable non-uniform magnetic configuration forms under a finite negative applied magnetic field, $H_n$, which can be associated to the usual "nucleation field" of micromagnetics equation. The field strength needs to be further increased for full reversal to occur, at $H_c > H_n$. For $t=1.2$ nm, the 300 K coercive field reaches values close to experimental values.

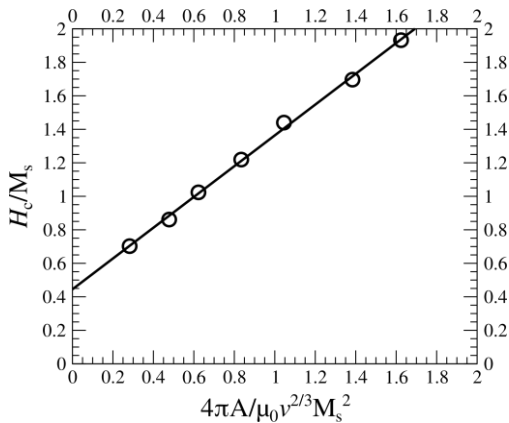

*Fig. 4 : $H_c/M_s$ vs. $4\pi A/\mu_0 v^{2/3} M_s^2$ derived from numerical modeling including thermal activation, for $Nd_2Fe_{14}B$ with 0.8 nm thick defect.*

To account for thermal activation, the minimum energy path during reversal is calculated using the nudged elastic band method [18]. The magnetic configuration of minimum energy at $H=H_c$ is $25k_BT$ lower than the configuration at the saddle point. The activation volume is calculated from the difference in magnetization between the saddle point and the local minimum.

We calculated the thermally-activated coercivity and associated activation volumes for the same cube model with a defect thickness of $t = 0.8$ nm, and material properties for $Nd_2Fe_{14}B$ ranging from T = 200 K to T = 500 K. Thermal activation reduced the coercivity value by between 31 and 47 percent, increasing with higher temperature. The associated activation volume at 200 K amounts to around 140 nm$^3$, to be compared to about 2200 nm$^3$ at 500 K. From the calculated temperature dependence of the coercive field, the calculated ratio $H_c/M_s$ is plotted in Figure 4 as a function of $4\pi A/\mu_0 v^{2/3} M_s^2$, with the results following a linear trend [19]. The derived slope is $\approx 1$, to be compared to $\alpha'_G = 2$ derived from Fig. 2. Despite the fact that the model appears to describe most experimental observations, this indicates that further model refining is needed.